\documentclass[twocolumn,showpacs,preprintnumbers,amsmath,amssymb]{revtex4}

\usepackage{graphicx}
\usepackage{dcolumn}
\usepackage{bm}

\begin{document}

\title{Metric Fluctuations from a NKK theory of gravity \\ in a de Sitter Expansion}

\author{Jos\'e Edgar Madriz Aguilar\footnote{E-mail address: edgar@ifm.umich.mx}}
 \affiliation{Instituto de F\'{\i}sica y Matem\'aticas, AP: 2-82, (58040) Universidad Michoacana de San Nicol\'as de  Hidalgo, 
Morelia, Michoac\'an, M\'exico.}

\author{Mauricio Bellini\footnote{E-mail address: mbellini@mdp.edu.ar}}
\affiliation{Departamento de F\'{\i}sica, Facultad de Ciencias Exactas y Naturales, Universidad Nacional de Mar del Plata, Funes 
3350, (7600) Mar del Plata, Argentina. Consejo Nacional de Ciencia y Tecnolog\'{\i}a.}


\begin{abstract}

En estas notas se presenta un formalismo recientemente introducido por los presentes autores para describir fluctuaciones 
escalares de la m\'etrica invariantes de norma en el contexto de una teor\'{\i}a de Kaluza-Klein no-compacta pentadimensional. En 
este an\'alisis se recupera uno de los resultados obtenidos t\'{\i}picamente bajo un tratamiento 4D semi-cl\'asico de inflaci\'on 
para las fluctuaciones en la densidad de energ\'{\i}a $\delta\rho /\rho \simeq 2\Phi$. Algo a resaltar es que el espectro para 
estas fluctuaciones es dependiente de la quinta coordenada. Este hecho nos permite establecer cotas para el n\'umero de onda 
asociado a la quinta dimensi\'on.\\

{\it Descriptores: fluctuaciones escalares de la m\'etrica, vac\'{\i}o aparente 5D, ecuaciones de Einstein linearizadas, quinta 
dimensi\'on no-compacta.}\\

A gauge invariant scalar metric fluctuations formalism from a Noncompact Kaluza-Klein (NKK) theory of gravity is presented in this 
talk notes. In this analysis we recover the well-known result $\delta\rho /\rho \simeq 2\Phi$ obtained typically in the standard 
4D semiclassical approach to inflation and also the spectrum of these fluctuations become dependent of the fifth (space-like) 
coordinate. This fact allows to establish an interval of values for the wave number associated with the fifth dimension.\\


\end{abstract}

\maketitle

\section{Introduction}

These talk notes are based on our recent work \cite{pertur1}. The goal is to study gauge invariant scalar metric fluctuations from 
a NKK theory of gravity in a de Sitter expansion. As we know the inflationary theory of the universe provides a physical mechanism 
to generate primordial energy density fluctuations. This is studied in the framework of the relativistic theory of cosmological 
perturbations. The theory of linearized gravitational perturbations in an expanding universe is used to describe the process of 
structure formation in the early universe \cite{riotto} and it is indispensable to relate inflationary scenarios with 
observational evidences mainly with the Cosmic Microwave Background (CMB) anisotropies. In the most widely accepted inflationary 
scenarios the dynamics is described by a quantum scalar field $\varphi$ named inflaton that is splitted into a homogeneous and an 
inhomogeneous components. Usually the homogeneous one is interpreted as a classical field  that drives the expansion, while the 
second one is responsible of the quantum fluctuations that originate the primordial energy density fluctuations \cite{linde90}.  
\\

On the other hand, physical theories in more than four dimensions have played an important role in mo\-dern physics including 
cosmology. The idea of extra dimensions in physics was proposed firstly by Gunnar Nordstr$\ddot{o}$m in 1914 \cite{Nor} and 
subsequently by Kaluza in 1921 \cite{Ka} and Klein in 1926 \cite{Kle}. They attempted to unify gravity with electromagnetism by 
introducing an extra dimension. Since then, the possible existence of extra dimensions got strong motivation and many interesting 
attempts to incorporate gravity and gauge interactions in an unique scheme have been made. Currently one of the theories with more 
impact in cosmology is the brane world scenario. In such framework the question about how large could extra dimensions be without 
getting into conflict with observational evidences, has a lot relevance. However for many researchers a more interesting question 
is how could this extra dimensions manifest themselves. According to brane world scenario matter should be localized onto an 
hypersurface (the brane) embedded in a higher dimensional space-time (the bulk) \cite{Abdel}. The main motivation of these models 
comes from string theories and their extension M-theory, which have suggested another approach to compactify extra dimensions. The 
proposal of great interest in cosmology is that our universe may be such a brane-like object where the standard model of particles 
is confined on a brane and only gravity and other exotic matter as some scalar fields (like the dilaton) can propagate in the bulk 
\cite{Phil}. \\

Another theory of great relevance and on which the present work is based is the Space-Time-Matter theory or Induced Matter theory. 
This theory can be thought as a noncompact Kaluza-Klein theory due to the fifth dimension is extended. In the 90's Paul Wesson, J. 
Ponce de Leon and collaborators, based in the Campbell-Magaard's theorem, showed that it is possible to interpret most  properties 
of matter in 4D as a result of the 5D Riemannian geometry. This formalism allows dependence on the fifth coordinate  and does not 
make assumptions about the topology of the fifth dimension. In other words, they proposed that 4D field equations with sources can 
be locally embedded in 5D field equations without sources \cite{WessonP}. The Campbell-Magaard's theorem establishes that any 
analytic N-dimensional Riemannian manifold can be locally embedded in a (N+1)-dimensional Ricci-flat manifold.
In the cosmological context there is a class of 5D cosmological models which are reduced to the usual 4D ones  by taking a 
foliation on the extra coordinate. These ideas will be implemented to develop the new formalism presented in this notes.\\

\section{Formalism}
We consider the action
\begin{equation} \label{action}
I=-\int d^{4}x \  d\psi\,\sqrt{\left|\frac{^{(5)}
\bar g}{^{(5)}\bar g_0}\right|} \ \left[
\frac{^{(5)}\bar R}{16\pi G}+ ^{(5)}{\cal L}(\varphi,\varphi_{,A})\right],
\end{equation}
for a scalar field $\varphi$, which is minimally coupled to gravity.
Since we are aimed to describe a manifold in apparent vacuum
the Lagrangian density ${\cal L}$ in (\ref{action}) should be only
kinetic in origin
\begin{equation}
^{(5)} {\cal L}(\varphi,\varphi_{,A}) = \frac{1}{2} g^{AB}
\varphi_{,A} \varphi_{,B},
\end{equation}
where $A,B$ can take the values 0,1,2,3,4 and the perturbed line element $dS^2=g_{AB} dx^A dx^B $ is given by
\begin{equation}\label{a}
dS^2 = \psi^2 \left(1+ 2 \Phi\right) dN^2 - \psi^2
\left(1- 2 \Psi\right) e^{2N} dr^2 - \left(1-Q\right) d\psi^2.
\end{equation}
Here, the fields $\Phi$, $\Psi$ and $Q$ are functions of the coordinates
[$N,\vec r(x,y,z),\psi$], where $N$, $x$, $y$, $z$ are dimensionless and the fifth coordinate $\psi$ has spatial units. Note that
$^{(5)}\bar R$ in the action (\ref{action})
is the Ricci scalar evaluated on the background metric
$\left(dS^2\right)_b = \bar g_{AB} dx^A dx^B$. In our case we
consider the background canonical metric
\begin{equation}\label{back}
\left(dS^2\right)_b
= \psi^2 dN^2 - \psi^2 e^{2N} dr^2
-d\psi^2,
\end{equation}
which is 3D spatially isotropic, homogeneous and flat\cite{otro}.
Moreover, the metric (\ref{back}) is globally flat (i.e.,
$\bar R^A_{BCD} =0$) and describes an apparent
vacuum $\bar G_{AB}=0$. The energy-momentum tensor is given by
\begin{equation}
T_{AB} = \varphi_{,A} \varphi_{,B}-
\frac{1}{2} g_{AB} \varphi_{,C} \varphi^{,C},
\end{equation}
which is obviously symmetric. Hence, using the fact that the
metric (\ref{a}) is also symmetric we obtain that $\Psi = \Phi$
and $Q=2\Phi$. Thus, the line element (\ref{a}) now becomes
\begin{equation}   \label{1}
dS^2
= \psi^2 \left(1+ 2 \Phi\right) dN^2 - \psi^2
\left(1- 2 \Phi\right) e^{2N} dr^2 - \left(1- 2 \Phi\right) d\psi^2,
\end{equation}
where the field $\Phi(N, \vec r, \psi)$ is the scalar metric perturbation
of the background 5D metric (\ref{back}).
For the metric (\ref{1}), $|^{(5)}\bar g|=\psi^8 e^{6N} $
is the absolute value of the determinant for
the background metric (\ref{back})
and
$|^{(5)} \bar g_0|=
\psi^8_0 e^{6N_0}$
is a dimensionalization constant, where $\psi_0$ and $N_0$ are constants.
Besides, $G=M^{-2}_p$ is the gravitational constant
and $M_p=1.2 \  10^{19} \  {\rm GeV}$ is the Planckian mass.
In this work we consider $N_0=0$, therefore
$\left|^{(5)}\bar g_0\right|=\psi^8_0$.
Here, the index $`` 0 "$ denotes the value
at the end of inflation.

On the other hand, the contravariant metric tensor, after
a $\Phi$-first order approximation, is given by
\begin{equation}     \label{3}
g^{NN}=\frac{(1-2\Phi)}{\psi^2},\,\, g^{ij}=-\frac{e^{-2N}
(1+2\Phi)}{\psi^2},\,\, g^{\psi\psi}=- (1+2\Phi),
\end{equation}
which can be written as $g^{AB} = \bar g^{AB} + \delta g^{AB}$,
being $\bar g^{AB}$ the contravariant background metric tensor and
$\delta g^{AB}$ their corresponding fluctuations. The dy\-na\-mics
for $\varphi$ and $\Phi$ are well described by the Lagrange and
Einstein equations, which we shall study in the following
subsections.

\subsection{5D Dynamics}

The Lagrange equations for the fields $\varphi$ and $\Phi$ are
res\-pec\-ti\-ve\-ly given by
\begin{eqnarray}
&& \frac{\partial^2\varphi}{\partial N^2}
+ 3\frac{\partial\varphi}{\partial N} - e^{-2N} \nabla^2_r \varphi -
\psi\left(\psi \frac{\partial^2\varphi}{\partial\psi^2}
+ 4 \frac{\partial\varphi}{\partial \psi}\right)
\nonumber \\
&-&
2\Phi \left[\frac{\partial^2\varphi}{\partial N^2}
+ 3\frac{\partial\varphi}{\partial N} - e^{-2N} \nabla^2_r \varphi
+ \psi\left(\psi \frac{\partial^2\varphi}{\partial\psi^2}
+ 4 \frac{\partial\varphi}{\partial\psi}\right)\right]\nonumber \\
&-& 2\left(\frac{\partial\varphi}{\partial N}
\frac{\partial\Phi}{\partial N} + \psi^2 \frac{\partial\Phi}{\partial\psi}
\frac{\partial\varphi}{\partial\psi} \right) =0,
\label{l1} \\
&& \left(\frac{\partial\varphi}{\partial N}\right)^2
+ e^{-2N} \left(\nabla\varphi\right)^2
+\psi^2 \left(\frac{\partial\varphi}{\partial\psi}\right)^2=0. \label{l2}
\end{eqnarray}
Now, we can make the semi classical approximation
$\varphi(N,\vec r, \psi) = \varphi_b(N,\psi) +
\delta\varphi(N,\vec r, \psi)$, such that $\varphi_b$
is the solution of eq. (\ref{l1})
in absence of the metric fluctuations [i.e.,for $\Phi =\delta\varphi =0$] and $\delta\varphi$ represents the quantum fluctuations 
of the inflaton field $\varphi$.
Hence, the Lagrange equations for $\varphi_b$ and
$\delta\varphi$ are
\begin{eqnarray}
&& \frac{\partial^2\varphi_b}{\partial N^2}
+ 3 \frac{\partial\varphi_b}{\partial N}
-\psi \left[ \psi \frac{\partial^2\varphi_b}{\partial\psi^2} +
4 \frac{\partial\varphi_b}{\partial\psi} \right]=0, \label{l3} \\
&& \frac{\partial^2\delta\varphi}{\partial N^2}
+ 3 \frac{\partial\delta\varphi}{\partial N}
- e^{-2N} \nabla^2_r \delta\varphi - \psi \left[
4\frac{\partial\delta\varphi}{\partial\psi}
+ \psi \frac{\partial^2\delta\varphi}{\partial\psi^2} \right] \nonumber \\
& - & 2 \frac{\partial\varphi_b}{\partial N} \frac{\partial\Phi}{\partial N}
-2 \psi^2 \left[ \frac{\partial\varphi_b}{\partial\psi}
\frac{\partial\Phi}{\partial\psi}
+\left( \frac{\partial^2\varphi_b}{\partial\psi^2}
+\frac{4}{\psi} \frac{\partial\varphi_b}{\partial\psi}\right) \Phi\right] =0.\nonumber \\
\label{l4}
\end{eqnarray}
Note that for $\Phi= \delta\varphi=0$, the eq. (\ref{l2})
transforms in
\begin{equation}\label{l5}
\left(\frac{\partial\varphi_b}{\partial N}\right)^2 + \psi^2
\left(\frac{\partial\varphi_b}{\partial\psi}\right)^2 =0,
\end{equation}
which will be useful later.\\

Considering the linearized field equations $\delta G_{AB} = -8\pi G
\delta T_{AB}$, after some algebra we reduce them to the form
\begin{eqnarray}
&&\frac{\partial^2 \Phi}{\partial N^2}+ 3 \frac{\partial\Phi}{\partial N}
- e^{-2N} \nabla^2_r \Phi - 2 \psi^2 \frac{\partial^2\Phi}{\partial\psi^2}
\nonumber \\
\label{eso}
&+& \frac{16 \pi G}{3} \Phi \left[
\left(\frac{\partial\varphi_b}{\partial N}\right)^2
+\psi^2 \left(\frac{\partial\varphi_b}{\partial\psi}\right)^2 \right] =0.
\end{eqnarray}
From eq. (\ref{l5}), the eq. (\ref{eso}) we obtain
\begin{equation} \label{phi}
\frac{\partial^2 \Phi}{\partial N^2}+ 3 \frac{\partial\Phi}{\partial N}
- e^{-2N} \nabla^2_r \Phi - 2 \psi^2 \frac{\partial^2\Phi}{\partial\psi^2}
=0,
\end{equation}
that is the equation of motion for the 5D scalar metric fluctuations
$\Phi(N,\vec r, \psi)$.

\section{Effective 4D de Sitter expansion}

In this section we study the effective 4D $\Phi$-dynamics in an effective
4D de Sitter background expansion of the universe, which is considered
3D (spatially) flat, isotropic and homogeneous.

\subsection{Ponce de Leon metric}

We consider the coordinate transformation \cite{plb2005}
\begin{equation}\label{trans}
t = \psi_0 N, \qquad R=\psi_0 r, \qquad \psi = \psi.
\end{equation}
Hence, the 5D background metric (\ref{back}) becomes
\begin{equation}\label{pdl}
\left(dS^2\right)_b = \left(\frac{\psi}{\psi_0}\right)^2
\left[dt^2 - e^{2t/\psi_0} dR^2\right]-d\psi^2,
\end{equation}
which is the Ponce de Leon metric\cite{pdlm}, that describes a 3D spatially
flat, isotropic and homogeneous extended (to 5D) Friedmann-Robertson-Walker
metric in a de Sitter expansion.
Here, $t$ is the cosmic time and $dR^2 = dX^2+dY^2+dZ^2$.
This Ponce de Leon metric is a special case of the separable models studied
by him, and is an example of the intensely studied class of canonical metrics
$dS^2 = \psi^2 g_{\mu\nu} dX^{\mu} dX^{\nu} - d\psi^2$ with $\mu ,\nu =0,1,2,3$ \cite{MLW}.
Now we can take a foliation $\psi=\psi_0$ in the metric (\ref{pdl}), such
that the effective 4D metric results
\begin{equation}\label{desitter}
\left(dS^2\right)_b \rightarrow \left(ds^2\right)_b
= dt^2 - e^{2t/\psi_0} dR^2,
\end{equation}
that describes a 4D expansion
of a 3D spatially flat, isotropic and homogeneous universe that
expands with a constant Hubble parameter $H=1/\psi_0$ and a 4D
scalar curvature $^{(4)} {\cal R} = 12 H^2$.
Hence, the effective 4D metric of (\ref{1}) on hypersurfaces $\psi=1/H$, is
\begin{equation}\label{4d}
dS^2 \rightarrow ds^2 = \left(1+2\Phi\right) dt^2 -
\left(1-2\Phi\right) e^{2Ht} dR^2.
\end{equation}
This metric describes the perturbed 4D de Sitter expansion of the universe,
where $\Phi(\vec R,t)$ is gauge-invariant.

\subsection{Dynamics of $\Phi$ in an effective 4D de Sitter expansion}

In order to study the 4D dynamics of the
gauge-invariant scalar metric fluctuations $\Phi(\vec R,t)$
in a background de Sitter expansion we transform the eq.(\ref{phi}) using the expressions (\ref{trans}) with the foliation 
$\psi=\psi_0=1/H$, eq.(\ref{phi}) acquires the form
\begin{equation}\label{mo4}
\frac{\partial^2\Phi}{\partial t^2} + 3H \frac{\partial \Phi}{\partial t}
-e^{-2 H t} \nabla^2_R \Phi - \left.
2 \frac{\partial^2\Phi}{\partial\psi^2}\right|_{\psi=H^{-1}}=0,
\end{equation}
where
$\left.
\frac{\partial^2\Phi}{\partial\psi^2}\right|_{\psi=H^{-1}}=
k^2_{\psi_0} \Phi$.
To simplify the structure of this equation we propose
the redefined quantum metric fluctuations
$\chi(\vec R,t) = e^{3H t/2} \Phi(\vec R, t)$,
thus eq.(\ref{mo4}) can be expressed in terms of $\chi$ as
\begin{equation}
\ddot\chi - e^{-2Ht} \nabla^2_R\chi - \left[
\frac{9}{4} H^2 + 4 k^2_{\psi_0}\right]\chi=0.
\end{equation}
Furthermore the redefined field $\chi(\vec R,t)$ can be expanded as
\begin{small}
\begin{equation}
\chi = \frac{1}{(2\pi)^{3/2}} \int d^3k_R
dk_{\psi} \left[ a_{k_R k_{\psi}} e^{i \vec k_R.\vec R}
\xi_{k_R k_{\psi}}(t) + cc\right] \delta(k_{\psi}-k_{\psi_0}),
\end{equation}
\end{small}
being $a_{k_Rk_{\psi}}$ and $a^{\dagger}_{k_R k_{\psi}}$ the annihilation and creation operators respectively, and $cc$ denoting 
the complex conjugate of the first term in brackets. These operators satisfy the commutator relations
\begin{displaymath}
\left[a_{k_Rk_{\psi}}, a^{\dagger}_{k'_{R}k'_{\psi}}\right] =
\delta^{(3)}\left(\vec k_R - \vec k'_R \right) \delta\left(
\vec k_{\psi} - \vec k'_{\psi}\right),
\end{displaymath}
\begin{displaymath}
\left[a_{k_Rk_{\psi}}, a_{k'_{R}k'_{\psi}}\right] =
\left[a^{\dagger}_{k_Rk_{\psi}}, a^{\dagger}_{k'_{R}k'_{\psi}}\right] =0.
\end{displaymath}
Hence, the dynamics of the time dependent modes $\xi_{k_R k_{\psi_0}}(t)$ is given
by
\begin{equation}
\ddot\xi_{k_R k_{\psi_0}}(t) + \left[k^2_R e^{-2Ht} - \left(\frac{9}{4} H^2+
4 k^2_{\psi_0}\right)\right]\xi_{k_R k_{\psi_0}}(t) =0,
\end{equation}
which has a general solution
\begin{equation}
\xi_{k_R k_{\psi_0}}(t) = A_1 {\cal H}^{(1)}_{\mu}[y(t)] +
A_2 {\cal H}^{(2)}_{\mu}[y(t)],
\end{equation}
where $\mu =\frac{1}{2}\sqrt{9+16 k^2_{\psi_0}/H^2}$
and $y(t) =\frac{1}{H}k_R e^{-H t}$.\\
Using the Bunch-Davies vacuum condition \cite{bd}, we obtain
\begin{equation}
\xi_{k_R k_{\psi_0}}(t) = i \sqrt{\frac{\pi}{4H}} {\cal H}^{(2)}_{\mu}[y(t)],
\end{equation}
which are the normalized time dependent modes of the field $\chi(\vec R,t)$.

\subsection{Energy density fluctuations}

In order to obtain the energy density fluctuations on the
effective 4D FRW metric, we calculate
\begin{equation}
\frac{\delta\rho}{\left<\rho\right>}
=
\left.\frac{\delta T^N_N}{\left<T^N_N\right>}
\right|_{t=\psi_0 N, R=\psi_0 r, \psi=1/H},
\end{equation}
being $\delta T_{NN}=-{1\over 2} \delta g_{NN} \varphi_{,L}
\varphi^{,L}$ linearized and where the brackets $<...>$
denote the expectation value on the 3D hypersurface $R(X,Y,Z)$.
Using the semiclassical expansion $\varphi(\vec R, t)=
\varphi_b (t) + \delta\varphi(\vec R,t)$,
after some algebra we obtain
\begin{small}
\begin{equation}
\frac{\delta\rho}{\left<\rho\right>} \simeq
2\Phi \left\{1-
\frac{\left<\left(\delta\dot\varphi\right)^2
+e^{-2Ht} \left(\nabla_R \delta\varphi\right)^2 +
2V(\delta\varphi)\right>}{\left(\dot\varphi_b\right)^2
+ 4 H^2 \left( \varphi_b\right)^2}\right\} \simeq 2\Phi,
\end{equation}
\end{small}
where we have considered the approximation
\begin{equation}\label{des}
\frac{\left<\left(\delta\dot\varphi\right)^2
+e^{-2Ht} \left(\nabla_R \delta\varphi\right)^2 +
2V(\delta\varphi)\right>}{\left(\dot\varphi_b\right)^2
+ 4 H^2 \left(\varphi_b\right)^2}
\ll 1,
\end{equation}
being $V(\delta\varphi)=V(\varphi) - V\left(\varphi_b\right)$
\begin{displaymath}
V(\delta\varphi) = -\frac{1}{2}\left[
\left. g^{\psi\psi} \left(\frac{\partial\varphi}{
\partial\psi}\right)^2 -
\bar g^{\psi\psi}
\left(\frac{\partial \varphi_b}{\partial
\psi}\right)^2\right]\right|_{\psi=H^{-1}},
\end{displaymath}
with $V(\varphi_b) = -\frac{1}{2} \bar g^{\psi\psi} \left.\left(\frac{\partial
\varphi_b}{\partial\psi}\right)^2\right|_{\psi=H^{-1}}=
2H^2 \left(\varphi_b\right)^2$.\\
It is important to notate that the approximation (\ref{des}) is
valid only during inflation on super Hubble scales (on the
infrared sector), on which the inflaton field fluctuations are
very ``smooth''. Finally, we can compute the amplitude of $\Phi
(\vec{R},t)$ for a de Sitter expansion on the infrared sector
($k_R \ll e^{Ht} H$) through the expression
\begin{equation}
\left<\Phi^2\right> = \frac{e^{-3Ht}}{(2\pi)^3}
{\Large\int}^{\epsilon e^{Ht} H}_0
d^3k_R \  \xi_{k_R} \xi^*_{k_R},
\end{equation}
where $\epsilon \simeq 10^{-3}$ is a dimensionless constant. Hence the squared
$\Phi$-fluctuations has a power-spectrum ${\cal P}(k_R)$ given by
\begin{equation}
{\cal P}(k_R) \sim k^{3-\sqrt{9+16k^2_{\psi_0}/H^2}}_R,
\end{equation}
which is nearly scale invariant for $k^2_{\psi_0}\psi^2_0=k^2_{\psi_0}/H^2 \ll 1$.
In other words, the 3D power-spectrum of the gauge-invariant metric
fluctuations depends of the wave number $k_{\psi_0}$
related to the fifth coordinate on the hypersurface $\psi=\psi_0\equiv
H^{-1}$. This 3D power spectrum corresponds to the spectral index
\begin{equation}
n_s=4-\sqrt{9+16 k^2_{\psi_0}/H^2} .
\end{equation}
On the other hand it is well known from observational data \cite{PDG} that
$n_s=0.97 \pm 0.03$. This fact allows to establish that $0 \le k_{\psi_0} < 0.15 \  H$, which is the main result of this paper.\\

\section{Final Comments}

In this notes, based on our recent work \cite{pertur1}, we have
studied 4D gauge-invariant me\-tric fluctuations from a NKK theory
of gravity. In particular we have exa\-mi\-ned these fluctuations
in an effective 4D de Sitter expansion for the universe using a
first-order expansion for the metric tensor. A very important
result of this formalism is the confirmation of the well known 4D
result $\delta\rho/\rho \simeq 2\Phi$ \cite{riotto}, during
inflation. Furthermore, the spectrum of the energy fluctuations
depends of the fifth coordinate. This fact allows to establish the
interval $k_{\psi_0} < 1.5\times 10^{-10} \  {\rm M_p}$, where we
have used the typical inflationary value $H=10^{-9} \ {\rm M_p}$.
Finally, an advantage of this formalism is that it could be
extended to other inflationary and cosmological models where the
expansion of the universe is governed by a single scalar field.

\bibliography{apssamp}

\end{document}